\batchmode
\makeatletter
\def\input@path{{"C:/Users/patilomkarsudhir/Desktop/Papers/ResNet Journal/Int J Adapt Control Signal Process/"}}
\makeatother
\RequirePackage{fixltx2e}
\pdfoutput=1
\documentclass[english]{IEEEtran}
\usepackage[T1]{fontenc}
\usepackage[latin9]{inputenc}
\usepackage{geometry}
\usepackage{color}
\usepackage{float}
\usepackage{amsmath}
\usepackage{amsthm}
\usepackage{amssymb}
\usepackage{stackrel}
\usepackage{graphicx}

\makeatletter

\providecommand{\tabularnewline}{\\}

\theoremstyle{definition}
\newtheorem{assumption}{Assumption}
\theoremstyle{remark}
\newtheorem{rem}{\protect\remarkname}
\theoremstyle{plain}
\newtheorem{thm}{\protect\theoremname}

\usepackage{cite}
\usepackage{multirow} 

\makeatother

\usepackage{babel}
\providecommand{\remarkname}{Remark}
\providecommand{\theoremname}{Theorem}

\begin{document}
\title{Lyapunov-Based Deep Residual Neural Network (ResNet) Adaptive Control}
\author{Omkar Sudhir Patil, Duc M. Le, Emily J. Griffis, and Warren E. Dixon\thanks{The authors are with the Department of Mechanical and Aerospace Engineering,
University of Florida, Gainesville FL 32611-6250 USA. Email: \{patilomkarsudhir,
ledan50, emilygriffis00, wdixon\}@ufl.edu.}\thanks{This research is supported in part by Office of Naval Research Grant
N00014-21-1-2481, AFOSR award number FA8651-21-F-1027, and FA9550-19-1-0169.
Any opinions, findings, and conclusions or recommendations expressed
in this material are those of the author(s) and do not necessarily
reflect the views of the sponsoring agency.}}
\maketitle
\begin{abstract}
Deep Neural Network (DNN)-based controllers have emerged as a tool
to compensate for unstructured uncertainties in nonlinear dynamical
systems. A recent breakthrough in the adaptive control literature
provides a Lyapunov-based approach to derive weight adaptation laws
for each layer of a fully-connected feedforward DNN-based adaptive
controller. However, deriving weight adaptation laws from a Lyapunov-based
analysis remains an open problem for deep residual neural networks
(ResNets). This paper provides the first result on Lyapunov-derived
weight adaptation for a ResNet-based adaptive controller. A nonsmooth
Lyapunov-based analysis is provided to guarantee asymptotic tracking
error convergence. Comparative Monte Carlo simulations are provided
to demonstrate the performance of the developed ResNet-based adaptive
controller. The ResNet-based adaptive controller shows a 64\% improvement
in the tracking and function approximation performance, in comparison
to a fully-connected DNN-based adaptive controller.
\end{abstract}

\section{Introduction}

Deep Neural Network (DNN)-based controllers have emerged as a tool
to compensate for unstructured uncertainties in nonlinear dynamical
systems. The success of DNN-based controllers is powered by the ability
of a DNN to approximate any continuous function over a compact domain
\cite{Kidger.Lyons2020}. A popular DNN-based control method is to
first perform a DNN-based offline system identification using sampled
input-output datasets that are collected by conducting experiments
\cite[Sec. 6.6]{Brunton.Kutz2019}. Then, using the identified DNN,
a feedforward term is constructed to compensate for uncertainty in
the system. However, the DNN weight estimates are not updated during
task execution, and hence, such an approach involves static implementation
of the DNN-based feedforward term. Since there is no continued learning
with most DNN methods, questions arise regarding how well the training
dataset matches the actual uncertainties in the system and the value
or quality of the static feedforward model. This strategy motivates
the desire for a large training dataset, but such data can be expensive
or impossible to obtain, including the need for higher order derivatives
that are typically not measurable. 

Unlike offline methods, a closed-loop adaptive feedforward term can
be developed by deriving real-time DNN weight adaptation laws from
a Lyapunov-based stability analysis. Various classical results \cite{Lewis1996a,Lewis1999a,Ge2002,Patre2008,Patre2010a}
use Lyapunov-based techniques to develop weight adaptation laws, but
only for single-hidden-layer networks. These results do not provide
weight adaptation laws for DNNs with more than one hidden layer, since
there are mathematical challenges posed by the nested and nonlinear
parameterization of a DNN that preclude the development of inner-layer
weight adaptation laws. Recent results such as \cite{Sun.Greene.ea2021,Joshi.Chowdhary2019,Joshi.Virdi.ea2020a}
develop Lyapunov-based adaptation laws for the output-layer weights
of a DNN. However, to update the inner-layer weights, results in \cite{Sun.Greene.ea2021,Joshi.Chowdhary2019,Joshi.Virdi.ea2020a}
collect datasets over discrete time-periods and iteratively identify
the inner-layer weights using offline training algorithms. To circumvent
offline identification of the inner-layer weights, the result in \cite{Le.Greene.ea2021}
provides a real-time inner-layer weight adaptation scheme based on
a modular approach. However, modular designs only offer constraints
on the adaptation laws and do not provide constructive insights on
designing the adaptation laws.

Our recent work in \cite{Patil.Le.ea2022} provides the first result
on Lyapunov-derived weight adaptation laws for each layer of a DNN-based
adaptive controller. To address the challenges posed by the nested
and nonlinearly-parameterized structure of the DNN, a recursive representation
of the DNN is developed. Then, a first-order Taylor series approximation
is recursively applied for each layer. Using a Lyapunov-based stability
analysis, the inner- and outer-layer weight adaptation laws are designed
to cancel coupling terms that result from the approximation strategy.
Although the result in \cite{Patil.Le.ea2022} provides Lyapunov-derived
weight adaptation laws for the DNN, the development is restricted
to fully-connected DNNs.

There are several limitations associated with standard DNN architectures
such as fully-connected and convolutional DNNs. Deeper networks typically
suffer from the problem of vanishing or exploding gradients, i.e.,
the rate of learning using a gradient-based update rule is highly
sensitive to the magnitude of DNN weights. Challenges faced from the
vanishing or exploding gradient problem are ubiquitous to both offline
training \cite{Goodfellow2016} and real-time weight adaptation \cite{Patil.Le.ea2022}.
Additionally, in applications such as image recognition, the performance
of a DNN is found to initially improve by increasing the depth of
the DNN. However, as the depth exceeds a threshold, performance rapidly
degrades \cite{He.Zhang.ea2016}.

To overcome the vanishing or exploding gradient problem and the degradation
of performance with the increasing depth of a DNN, results in \cite{He.Zhang.ea2016}
introduce shortcut connections across layers, i.e., a feedforward
connection between layers that are separated by more than one layer.
DNNs with a shortcut connection are known as deep residual neural
networks (ResNets). Offline results in \cite{Hardt.Ma2017} and \cite{Nar.Sastry2018}
offer mathematical explanations for why ResNets perform better than
non-residual DNNs. In \cite{Hardt.Ma2017}, the parameterization of
a non-residual DNN is shown to cause difficulties in training DNN
layers to approximate the identity function. As explained in \cite{Hardt.Ma2017},
for a DNN to achieve a good training accuracy, the DNN layers must
be able to approximate the identity function well. Since a shortcut
connection in ResNets is represented using an identity function, ResNets
provide an improved performance when compared to non-residual DNNs.
Additionally, the result in \cite{Nar.Sastry2018} provides explanations
from Lyapunov stability theory on why ResNets are easier to train
offline using the gradient descent algorithm as compared to non-residual
DNNs. The shortcut connections in ResNets facilitate the stability
of the equilibria of gradient descent dynamics for a larger set of
step sizes or initial weights as compared to non-residual DNNs.

Although there has been significant research across various applications
involving ResNets \cite{He.Zhang.ea2016,Tai.Yang.ea2017,Li.Fang.ea2018,Boroumand.Chen.ea2018,Tan.Qian.ea2018},
the approximation power of ResNets has not yet been explored for adaptive
control problems. Developing a ResNet-based adaptive feedforward control
term with real-time weight adaptation laws is an open problem. Although
real-time weight adaptation laws are developed for fully-connected
feedforward DNNs in \cite{Patil.Le.ea2022}, the shortcut connections
in ResNets pose additional mathematical challenges. Unlike fully-connected
DNNs, the shortcut connection prevents a recursive application of
Taylor series approximation for each layer of the ResNet. As a result,
it is difficult to generate the coupling terms that are generated
using the approximation strategy in \cite{Patil.Le.ea2022}, that
can be canceled using the weight adaptation laws in the Lyapunov-based
analysis.

Our preliminary work in \cite{Patil.Le.ea.2022} and this paper provide
the first result on Lyapunov-derived adaptation laws for the weights
of each layer of a ResNet-based adaptive controller for uncertain
nonlinear systems. To overcome the mathematical challenges posed by
the residual network architecture, the ResNet is expressed as a composition
of building blocks that involve a shortcut connection across a fully-connected
DNN. Then, a constructive Lyapunov-based approach is provided to derive
weight adaptation laws for the ResNet using the gradient of each DNN
building block. A nonsmooth Lyapunov-based analysis is provided to
guarantee asymptotic tracking error convergence. Unlike our preliminary
work in \cite{Patil.Le.ea.2022}, which involved a ResNet with only
one shortcut connection, this paper provides weight adaptation laws
for a general ResNet that has an arbitrary number of shortcut connections.
The development of adaptation laws for ResNets with an arbitrary number
of shortcut connections is challenging due to the complexity of the
architecture. This challenge is addressed by constructing a recursive
representation of the ResNet which involves a composition of an arbitrary
number of building blocks. Then, based on the recursive representation
of the ResNet architecture, a first-order Taylor series approximation
is applied, which is then utilized to yield the Lyapunov-based adaptation
laws. Additionally, unlike our preliminary work in \cite{Patil.Le.ea.2022}
which did not provide simulations, this paper provides comparative
Monte Carlo simulations to demonstrate the performance of the developed
ResNet-based adaptive controller, and the results are compared with
an equivalent fully-connected DNN-based adaptive controller \cite{Patil.Le.ea2022}.
Since the performance of ResNet and DNN-based adaptive controllers
is sensitive to weight initialization, the Monte Carlo approach is
used to provide a fair comparison between the two architectures. In
the Monte Carlo comparison, 10,000 simulations are performed, where
the initial weights in each simulation are selected from a uniform
random distribution, and a cost function is evaluated for each simulation.
Then, the simulation results yielding the least cost for both architectures
are compared. The ResNet-based adaptive controller shows a 64\% improvement
in the tracking and function approximation performance, in comparison
to a fully-connected DNN-based adaptive controller.

\subsubsection*{Notation and Preliminaries}

The space of essentially bounded Lebesgue measurable functions is
denoted by $\mathcal{L}_{\infty}$. The right-to-left matrix product
operator is represented by $\stackrel{\curvearrowleft}{\prod}$, i.e.,
$\stackrel{\curvearrowleft}{\stackrel[p=1]{m}{\prod}}A_{p}=A_{m}\ldots A_{2}A_{1}$
and $\stackrel{\curvearrowleft}{\stackrel[p=a]{m}{\prod}}A_{p}=I$
if $a>m$. The Kronecker product is denoted by $\otimes$. Function
compositions are denoted using the symbol $\circ$, e.g., $(g\circ h)(x)=g(h(x))$,
given suitable functions $g$ and $h$. The Filippov set-valued map
defined in \cite[Equation 2b]{Paden1987} is denoted by $K\left[\cdot\right]$.
The notation $\overset{a.a.t.}{(\cdot)}$ denotes that the relation
$(\cdot)$ holds for almost all time (a.a.t.). Consider a Lebesgue
measurable and locally essentially bounded function $h:\mathbb{R}^{n}\times\mathbb{R}_{\geq0}\to\mathbb{R}^{n}$.
Then, the function $y:\mathcal{I}\to\mathbb{R}^{n}$ is called a Filippov
solution of $\dot{y}=h(y,t)$ on the interval $\mathcal{I}\subseteq\mathbb{R}_{\geq0}$
if $y$ is absolutely continuous on $\mathcal{I}$ and $\dot{y}\overset{a.a.t.}{\in}K\left[h\right](y,t)$.
Given $w\in\mathbb{R}$ and some functions $f$ and $g$, the notation
$f(w)=\mathcal{O}^{m}(g(w))$ means that there exists some constants
$M\in\mathbb{R}_{>0}$ and $w_{0}\in\mathbb{R}$ such that $\left\Vert f(w)\right\Vert \leq M\left\Vert g(w)\right\Vert ^{m}$
for all $w\geq w_{0}$. Given some matrix $A\triangleq\left[a_{i,j}\right]\in\mathbb{R}^{n\times m}$,
where $a_{i,j}$ denotes the element in the $i^{th}$ row and $j^{th}$
column of $A$, the vectorization operator is defined as $\mathrm{vec}(A)\triangleq[a_{1,1},\ldots,a_{1,m},\ldots,a_{n,1},\ldots,a_{n,m}]^{T}\in\mathbb{R}^{nm}$.
The $p$-norm is denoted by $\left\Vert \cdot\right\Vert _{p}$, where
the subscript is suppressed when $p=2$. The Frobenius norm is denoted
by $\left\Vert \cdot\right\Vert _{F}\triangleq\left\Vert \mathrm{vec}(\cdot)\right\Vert $.
Given any $A\in\mathbb{R}^{p\times a}$, $B\in\mathbb{R}^{a\times r}$,
and $C\in\mathbb{R}^{r\times s}$, the vectorization operator satisfies
the property \cite[Proposition 7.1.9]{Bernstein2009} 
\begin{eqnarray}
\mathrm{vec}(ABC) & = & (C^{T}\otimes A)\mathrm{vec}\left(B\right).\label{eq:vec_prop}
\end{eqnarray}
Differentiating (\ref{eq:vec_prop}) on both sides with respect to
$\mathrm{vec}\left(B\right)$ yields the property
\begin{eqnarray}
\frac{\partial}{\partial\mathrm{vec}\left(B\right)}\mathrm{vec}(ABC) & = & (C^{T}\otimes A).\label{eq:vec_diff_prop}
\end{eqnarray}

\section{Unknown System Dynamics and Control Design}

Consider a control-affine nonlinear dynamic system modeled as
\begin{eqnarray}
\dot{x} & = & f(x)+u,\label{eq: xDot}
\end{eqnarray}
where $x:\mathbb{R}_{\ge0}\rightarrow\mathbb{R}^{n}$ denotes a Filippov
solution to (\ref{eq: xDot}), $f:\mathbb{R}^{n}\rightarrow\mathbb{R}^{n}$
denotes an unknown differentiable drift vector field, and $u:\mathbb{R}_{\ge0}\rightarrow\mathbb{R}^{n}$
denotes a control input.\footnote{The control effectiveness term is omitted to better focus on the specific
contributions of this paper without loss of generality. The method
in \cite{Sun.Greene.ea2021} can be used with the developed method
in the case where the system involves an uncertain control effectiveness
term.} Let the tracking error $e:\mathbb{R}_{\ge0}\rightarrow\mathbb{R}^{n}$
be defined as
\begin{eqnarray}
e & \triangleq & x-x_{d},\label{eq: e}
\end{eqnarray}
where $x_{d}:\mathbb{R}_{\ge0}\rightarrow\mathbb{R}^{n}$ denotes
a continuously differentiable reference trajectory. The reference
trajectory is designed such that $\left\Vert x_{d}(t)\right\Vert \leq\overline{x_{d}}\;\forall t\in\mathbb{R}_{\ge0}$
and $\dot{x}_{d}\in\mathcal{L}_{\infty},$ where $\overline{x_{d}}\in\mathbb{R}_{>0}$
is a constant. The control objective is to design a ResNet-based adaptive
controller that achieves asymptotic tracking error convergence.

\subsection{ResNet Architecture}

\begin{figure}
\begin{centering}
\includegraphics[width=8cm]{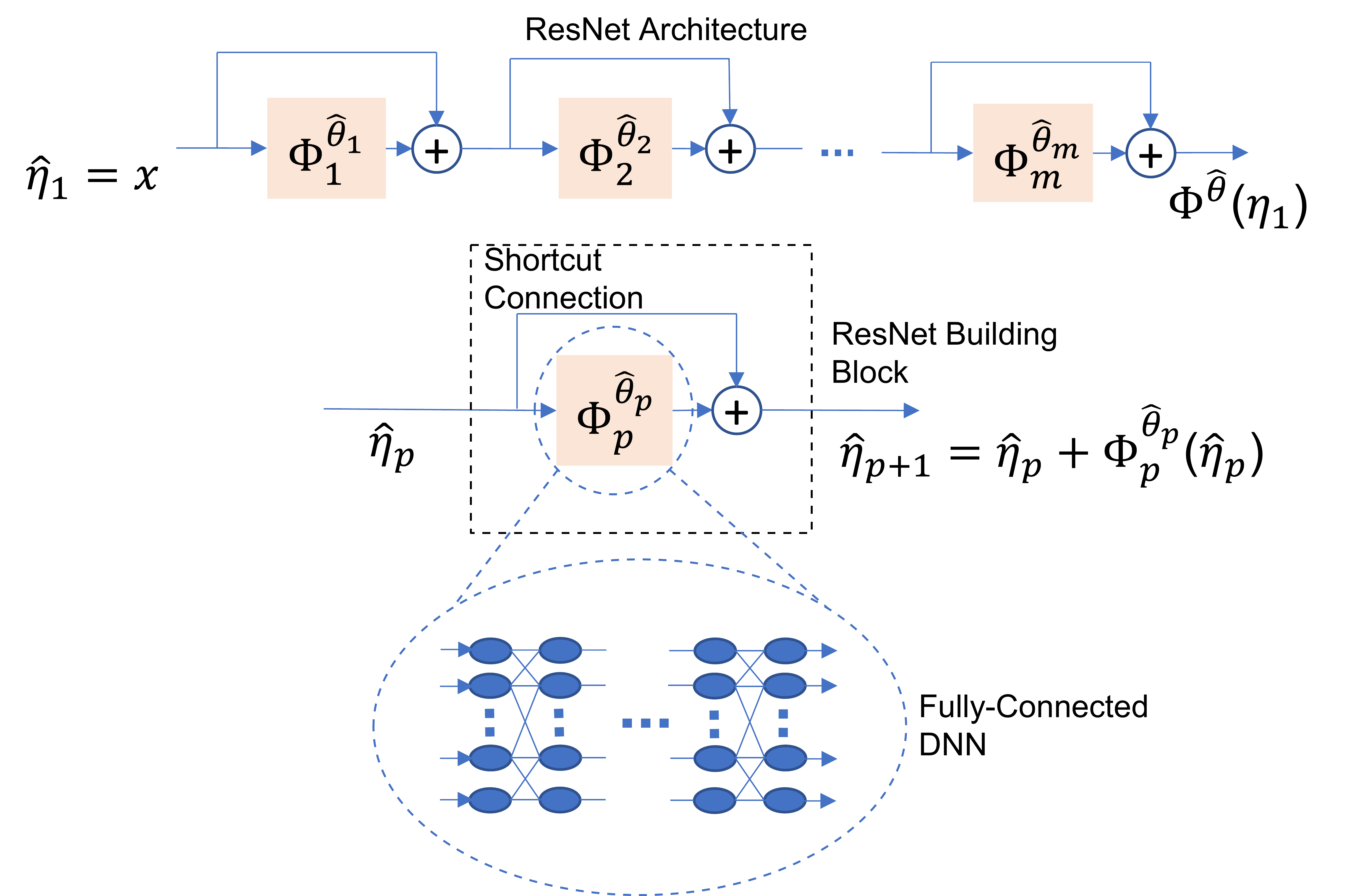}
\par\end{centering}
\centering{}\caption{\label{fig:ResNet1} Illustration of the ResNet architecture in (\ref{eq:ResNet1_arch}).
The ResNet is shown at the top of the figure and is composed of building
blocks that involve a shortcut connection across a fully-connected
DNN component. The fully-connected DNN component for the $p^{th}$
building block (bottom) is denoted by $\Phi_{p}^{\theta_{p}}$ for
all $p\in\{1,\ldots,m\}$, where the input and the vector of weights
of $\Phi_{p}$ are denoted by $\eta_{p}$ and $\theta_{p}$, respectively.
Then the output of the $p^{th}$ building block after considering
the shortcut connection is represented by $\eta_{p+1}=\eta_{p}+\Phi_{p}^{\theta_{p}}(\eta_{p})$
for all $p\in\{1,\ldots,m-1\}$, and the output of the ResNet is $\eta_{m}+\Phi_{m}^{\theta_{m}}\left(\eta_{m}\right)$.}
\end{figure}
The unknown drift vector field $f$ can be approximated using a ResNet.
A ResNet is modeled using building blocks that involve a shortcut
connection across a fully-connected DNN \cite{He.Zhang.ea2016}. Let
$\Phi_{p}:\mathbb{R}^{L_{p,0}}\times\mathbb{R}^{L_{p,0}\times L_{p,1}}\times\ldots\times\mathbb{R}^{L_{p,k_{p}}\times L_{p,k_{p}+1}}\to\mathbb{R}^{L_{p,k_{p}+1}}$
denote the $p^{th}$ fully-connected DNN block defined as $\Phi_{p}(\eta_{p},V_{p,0},\ldots,V_{p,k_{p}})\triangleq\left(V_{p,k_{p}}^{T}\phi_{p,k_{p}}\circ...\circ V_{p,1}^{T}\phi_{p,1}\right)\left(V_{p,0}^{T}\eta_{p}\right)$
for all $p\in\{1,\ldots,m\}$, where $\eta_{p}\in\mathbb{R}^{L_{p,0}}$
denotes the input of $\Phi_{p}$, $k_{p}\in\mathbb{Z}_{>0}$ denotes
the number of hidden layers in $\Phi_{p}$, and $m\in\mathbb{Z}_{>0}$
denotes the number of building blocks. Additionally, $L_{p,j}\in\mathbb{Z}_{>0}$
denotes the number of nodes, and $V_{p,j}\in\mathbb{R}^{L_{p,j}\times L_{p,j+1}}$
denotes the weight matrix in the $j^{th}$ layer of $\Phi_{p}$ for
all $(p,j)\in\{1,\ldots,m\}\times\{0,\ldots,k_{p}\}$. Similarly,
$\phi_{p,j}:\mathbb{R}^{L_{p,j}}\to\mathbb{R}^{L_{p,j}}$ denotes
a vector of smooth activation functions.\footnote{For the case of DNNs with nonsmooth activation functions (e.g., rectified
linear unit (ReLU), leaky ReLU, maxout etc.), the reader is referred
to \cite{Patil.Le.ea2022} where a switched analysis is provided to
account for the nonsmooth nature of activation functions. To better
focus on our main contribution without loss of generality, we restrict
our attention to smooth activation functions.} If the ResNet involves multiple types of activation functions at
each layer, then $\phi_{p,j}$ may be represented as $\phi_{p,j}\triangleq\left[\varsigma_{p,j,1}\begin{array}{cc}
\ldots & \varsigma_{p,j,L_{p,j}}\end{array}\right]^{T}$, where $\varsigma_{p,j,i}:\mathbb{R}\to\mathbb{R}$ denotes the activation
function at the $i^{\mathrm{th}}$ node of the $j^{\mathrm{th}}$
layer of $\Phi_{p}$.\footnote{Bias terms are omitted for simplicity of the notation.}
All the weights of $\Phi_{p}$ can be represented by the vector $\theta_{p}\triangleq\left[\begin{array}{ccc}
\mathrm{vec}(V_{p,0})^{T} & \ldots & \mathrm{vec}(V_{p,k_{p}})^{T}\end{array}\right]^{T}\in\mathbb{R}^{\Sigma_{j=0}^{k_{p}}L_{p,j}L_{p,j+1}}$. The fully-connected block $\Phi_{p}$ can be expressed as $\Phi_{p}=V_{p,k_{p}}^{T}\varphi_{p,k_{p}}$,
where $\varphi_{p,0}:\mathbb{R}^{L_{p,0}}\to\mathbb{R}^{L_{p,0}}$
and $\varphi_{p,j}:\mathbb{R}^{L_{p,0}}\times\mathbb{R}^{L_{p,0}\times L_{p,1}}\times\ldots\times\mathbb{R}^{L_{p,j-1}\times L_{p,j}}\to\mathbb{R}^{L_{p,j}}$
$\forall j\in\{1,\ldots,k_{p}\}$ denote the recursive relation defined
as
\begin{eqnarray}
\varphi_{p,j} & \triangleq & \begin{cases}
\phi_{p,j}\left(V_{p,j-1}^{T}\varphi_{p,j-1}\right), & j\in\left\{ 1,\ldots,k_{p}\right\} ,\\
\eta_{p}, & j=0.
\end{cases}\label{eq:varphi_pj}
\end{eqnarray}
The arguments of $\varphi_{p,j}$ are suppressed for notational brevity.
Let $\phi_{p,j}^{\prime}:\mathbb{R}^{L_{p,j}}\to\mathbb{R}^{L_{p,j}\times L_{p,j}}$
be defined as $\phi_{p,j}^{\prime}(y)\triangleq\frac{\partial}{\partial y}\phi_{p,j}(y)$
$\forall y\in\mathbb{R}^{L_{p,j}}$. The short-hand notation $\Phi_{p}^{\theta_{p}}(\eta_{p})\triangleq\Phi_{p}(\eta_{p},V_{p,0},V_{p,1},\ldots,V_{p,k_{p}})$
 is defined for notational brevity in the subsequent development.
Then the output of the $p^{th}$ building block is given by $\eta_{p}+\Phi_{p}^{\theta_{p}}(\eta_{p})$,
where the addition of the input term $\eta_{p}$ represents the shortcut
connection across $\Phi_{p}$. As shown in Figure \ref{fig:ResNet1},
the ResNet $\Phi:\mathbb{R}^{n}\times\mathbb{R}^{\Sigma_{p=1}^{m}\Sigma_{j=0}^{k_{p}}L_{p,j}L_{p,j+1}}\to\mathbb{R}^{n}$,
which defines the mapping $(\eta_{1},\theta)\mapsto\Phi^{\theta}\left(\eta_{1}\right)$,
is modeled as \cite{He.Zhang.ea2016}
\begin{eqnarray}
\Phi^{\theta}\left(\eta_{1}\right) & \triangleq & \eta_{m}+\Phi_{m}^{\theta_{m}}\left(\eta_{m}\right),\label{eq:ResNet1_arch}
\end{eqnarray}
where $\theta\triangleq\left[\begin{array}{ccc}
\theta_{1}^{T} & \ldots & \theta_{m}^{T}\end{array}\right]^{T}\in\mathbb{R}^{\Sigma_{p=1}^{m}\Sigma_{j=0}^{k_{p}}L_{p,j}L_{p,j+1}}$ denotes the vector of weights for the entire ResNet, and $\eta_{m}$
is evaluated using the recursive relation
\begin{eqnarray}
\eta_{p} & = & \begin{cases}
\eta_{p-1}+\Phi_{p-1}^{\theta_{p-1}}\left(\eta_{p-1}\right), & p\in\{2,\ldots,m\},\\
x, & p=1.
\end{cases}\label{eq:eta_p_recursion}
\end{eqnarray}
The recursive relation in (\ref{eq:eta_p_recursion}) has valid dimensions
under the constraint $L_{1,0}=L_{1,k_{1}+1}=L_{2,0}=L_{2,k_{2}+1}=\ldots=L_{m,0}=L_{m,k_{m}+1}=n$.
To facilitate the subsequent development, the following assumption
is made.
\begin{assumption}
\label{assm:UAP} The function space of ResNets given by (\ref{eq:ResNet1_arch})
is dense in $\mathcal{C}(\Omega)$ with respect to the supremum norm,
where $\mathcal{C}(\Omega)$ denotes the space of functions continuous
over the compact set $\Omega\subset\mathbb{R}^{n}$.
\end{assumption}
\begin{rem}
\label{rem:UAP} Assumption \ref{assm:UAP} implies that ResNets satisfy
the universal function approximation property that is well-known for
various DNN architectures \cite{Kidger.Lyons2020}. The universal
function approximation property of ResNets is a common assumption
that is widely used in the deep learning literature, and has been
rigorously established for ResNets with specific activation functions
in \cite{Lin.Jegelka2018} and \cite{Tabuada.Gharesifard2020a}. 
\end{rem}
Consider any vector field $f\in\mathcal{C}(\Omega)$ and a prescribed
accuracy $\overline{\varepsilon}\in\mathbb{R}_{>0}$. Then by Assumption
\ref{assm:UAP}, there exists a ResNet $\Phi$ with sufficiently large
$m,k_{p},L_{p,j}$ $\forall(p,j)\in\{1,\ldots,m\}\times\{0,\ldots,k_{p}\}$
and a corresponding vector of ideal weights $\theta^{*}\in\mathbb{R}^{\Sigma_{p=1}^{m}\Sigma_{j=0}^{k_{p}}L_{p,j}L_{p,j+1}}$
such that $\sup_{x\in\Omega}\left\Vert f(x)-\Phi^{\theta^{*}}(x)\right\Vert \leq\overline{\varepsilon}$.
Therefore, the drift vector field $x\mapsto f(x)$ can be modeled
as\footnote{If $f$ has different input-output dimensions, i.e., $f:\mathbb{R}^{\mu}\to\mathbb{R}^{n}$
with the input dimension $\mu$, the ResNet architecture can be modified
with an extra fully-connected DNN block $\Phi_{0}^{\theta_{0}}(x)\in\text{\ensuremath{\mathbb{R}^{\mu}}}$
at the input to account for the difference in input and output domains.}
\begin{eqnarray}
f(x) & = & \Phi^{\theta^{*}}(x)+\varepsilon(x),\label{eq:UAP}
\end{eqnarray}
when $x\in\Omega$, where $\varepsilon:\mathbb{R}^{n}\rightarrow\mathbb{R}^{n}$
denotes an unknown function reconstruction error that can be bounded
as $\sup_{x\in\Omega}\left\Vert \varepsilon(x)\right\Vert \leq\overline{\varepsilon}$.

To facilitate the subsequent analysis, the following assumption is
made (cf., \cite[Assumption 1]{Lewis1996b}).
\begin{assumption}
\label{assm:theta bound} There exists a known constant $\overline{\theta}\in\mathbb{R}_{>0}$
such that the unknown ideal ResNet weights can be bounded as $\left\Vert \theta^{*}\right\Vert \leq\overline{\theta}$.
\end{assumption}

\subsection{Control and Adaptation Laws}

The ResNet-based model in (\ref{eq:UAP}) can be leveraged to approximate
the unknown drift vector field $f$. However, since the ideal weights
are unknown, adaptive weight estimates are developed. The adaptive
weight estimate for the $j^{th}$ layer of $\Phi_{p}$ is denoted
by $\hat{V}_{p,j}:\mathbb{R}_{\ge0}\rightarrow\mathbb{R}^{L_{p,j}\times L_{p,j+1}}$
$\forall(p,j)\in\{1,\ldots,m\}\times\{0,\ldots,k_{p}\}$. The weight
estimate for the $p^{th}$ building block $\hat{\theta}_{p}:\mathbb{R}_{\ge0}\rightarrow\mathbb{R}^{\Sigma_{j=0}^{k_{p}}L_{p,j}L_{p,j+1}}$
is defined as $\hat{\theta}_{p}=\left[\mathrm{vec}(\hat{V}_{p,0})^{T},\ldots,\mathrm{vec}(\hat{V}_{p,k_{p}})^{T}\right]^{T}$
for all $p\in\{1,\ldots,m\}$, the weight estimate for the ResNet
$\hat{\theta}:\mathbb{R}_{\ge0}\rightarrow\mathbb{R}^{\Sigma_{p=1}^{m}\Sigma_{j=0}^{k_{p}}L_{p,j}L_{p,j+1}}$
is defined as $\hat{\theta}\triangleq\left[\begin{array}{ccc}
\hat{\theta}_{1}^{T} & \ldots & \hat{\theta}_{m}^{T}\end{array}\right]^{T}$, and the ResNet-based adaptive estimate of $f(x)$ $\forall x\in\Omega$
is denoted by $\Phi^{\hat{\theta}}(x)$. The weight estimation error
$\tilde{\theta}:\mathbb{R}_{\ge0}\rightarrow\mathbb{R}^{\Sigma_{p=1}^{m}\Sigma_{j=0}^{k_{p}}L_{p,j}L_{p,j+1}}$
is defined as $\tilde{\theta}\triangleq\theta^{*}-\hat{\theta}$.
Based on the subsequent stability analysis, the adaptation law for
the weight estimates of the ResNet in (\ref{eq:ResNet1_arch}) is
designed as
\begin{eqnarray}
\dot{\hat{\theta}} & \triangleq & \Gamma\Phi^{\prime T}e,\label{eq:adaptation_law}
\end{eqnarray}
where $\Gamma\in\mathbb{R}^{\Sigma_{p=1}^{m}\Sigma_{j=0}^{k_{p}}L_{p,j}L_{p,j+1}\times\Sigma_{p=1}^{m}\Sigma_{j=0}^{k_{p}}L_{p,j}L_{p,j+1}}$
denotes a positive-definite adaptation gain matrix, and $\Phi^{\prime}\in\mathbb{R}^{n\times\Sigma_{p=1}^{m}\Sigma_{j=0}^{k_{p}}L_{p,j}L_{p,j+1}}$
is a short-hand notation denoting the  $\Phi^{\prime}\triangleq\frac{\partial\Phi^{\hat{\theta}}(x)}{\partial\hat{\theta}}$.
The term $\Phi^{\prime}$ can be evaluated as follows. Let $\hat{\eta}_{p}\in\mathbb{R}^{L_{p,0}}$
be defined as
\begin{equation}
\hat{\eta}_{p}=\begin{cases}
x, & p=1,\\
\hat{\eta}_{p-1}+\Phi_{p-1}^{\hat{\theta}_{p-1}}\left(\hat{\eta}_{p-1}\right), & p\in\{2,\ldots,m\}.
\end{cases}\label{eq:eta_hat_p_recursion}
\end{equation}
Then, it follows that $\Phi^{\hat{\theta}}\left(x\right)=\hat{\eta}_{m}+\Phi_{m}^{\hat{\theta}_{m}}(\hat{\eta}_{m})$.
To facilitate the subsequent development, the short-hand notations
$\Phi_{p}^{\prime}\triangleq\left(\frac{\partial\Phi^{\hat{\theta}}(x)}{\partial\hat{\theta}_{p}}\right)$,
$\Lambda_{p}\triangleq\frac{\partial\Phi_{p}^{\hat{\theta}_{p}}\left(\hat{\eta}_{p}\right)}{\partial\hat{\theta}_{p}}$,
$\Lambda_{p,j}\triangleq\frac{\partial\Phi_{p}^{\hat{\theta}_{p}}\left(\hat{\eta}_{p}\right)}{\partial\mathrm{vec}(\hat{V}_{p,j})}$,
and $\Xi_{p}\triangleq\frac{\partial\Phi_{p}^{\hat{\theta}_{p}}\left(\hat{\eta}_{p}\right)}{\partial\hat{\eta}_{p}}$
are introduced. Then $\Phi^{\prime}=\left[\left(\frac{\partial\Phi^{\hat{\theta}}(x)}{\partial\hat{\theta}_{1}}\right),\ldots,\left(\frac{\partial\Phi^{\hat{\theta}}(x)}{\partial\hat{\theta}_{m}}\right)\right]$
can be expressed as
\begin{eqnarray}
\Phi^{\prime} & \triangleq & \left[\begin{array}{ccc}
\Phi_{1}^{\prime}, & \ldots, & \Phi_{m}^{\prime}\end{array}\right].\label{eq:Phi_prime}
\end{eqnarray}
Using the chain rule, the term $\Phi_{p}^{\prime}$ can be computed
as
\begin{equation}
\Phi_{p}^{\prime}=\left(\stackrel{\curvearrowleft}{\stackrel[l=p+1]{m}{\prod}}\left(I_{n}+\Xi_{l}\right)\right)\Lambda_{p},\,\forall p\in\{1,\ldots,m\}.\label{eq:Phi_p_prime}
\end{equation}
In (\ref{eq:Phi_p_prime}), the terms $\Lambda_{p}$ and $\Xi_{p}$,
for all $p\in\{1,\ldots,m\}$, can be computed as follows. Since $\hat{\theta}_{p}=\left[\mathrm{vec}(\hat{V}_{p,0})^{T},\ldots,\mathrm{vec}(\hat{V}_{p,k_{p}})^{T}\right]^{T}$,
it follows that $\frac{\partial\Phi_{p}^{\hat{\theta}_{p}}\left(\hat{\eta}_{p}\right)}{\partial\hat{\theta}_{p}}=\left[\left(\frac{\partial\Phi_{p}^{\hat{\theta}_{p}}\left(\hat{\eta}_{p}\right)}{\partial\mathrm{vec}(\hat{V}_{p,0})}\right),\ldots,\left(\frac{\partial\Phi_{p}^{\hat{\theta}_{p}}\left(\hat{\eta}_{p}\right)}{\partial\mathrm{vec}(\hat{V}_{p,k_{p}})}\right)\right]$.
Therefore, using the definitions of $\Lambda_{p}$ and $\Lambda_{p,j}$
yields
\begin{equation}
\Lambda_{p}=\left[\begin{array}{cccc}
\Lambda_{p,0} & \Lambda_{p,1} & \ldots & \Lambda_{p,k_{p}}\end{array}\right],\forall p\in\{1,\ldots,m\}.\label{eq:Lambda_p}
\end{equation}
For brevity in the subsequent development, the short-hand notations
$\hat{\varphi}_{p,j}\triangleq\varphi_{p,j}(\hat{\eta}_{p},\hat{V}_{p,0},\ldots,\hat{V}_{p,j})$
and $\hat{\varphi}_{p,j}^{\prime}\triangleq\varphi_{p,j}^{\prime}(\hat{\eta}_{p},\hat{V}_{p,0},\ldots,\hat{V}_{p,j})$
are introduced. Using (\ref{eq:varphi_pj}), the chain rule, and the
property of vectorization operators in (\ref{eq:vec_diff_prop}),
the terms $\Lambda_{p,0}$ and $\Lambda_{p,j}$ in (\ref{eq:Lambda_p})
can be computed as
\begin{eqnarray}
\Lambda_{p,0} & = & \left(\stackrel{\curvearrowleft}{\stackrel[l=1]{k_{p}}{\prod}}\hat{V}_{p,l}^{T}\hat{\varphi}_{p,l}^{\prime}\right)(I_{L_{p,1}}\otimes\hat{\eta}_{p}^{T}),\label{eq:Lambda_p0}
\end{eqnarray}
and
\begin{eqnarray}
\Lambda_{p,j} & = & \left(\stackrel{\curvearrowleft}{\stackrel[l=j+1]{k_{p}}{\prod}}\hat{V}_{p,l}^{T}\hat{\varphi}_{p,l}^{\prime}\right)(I_{L_{p,j+1}}\otimes\hat{\varphi}_{p,j}^{T}),\label{eq:Lambda_pj}
\end{eqnarray}
for all $(p,j)\in\{1,\ldots,m\}\times\{1,\ldots,k_{p}\}$, respectively.
Similarly, the term $\Xi_{p}$ can be computed as
\begin{eqnarray}
\Xi_{p} & = & \left(\stackrel{\curvearrowleft}{\stackrel[l=1]{k_{p}}{\prod}}\hat{V}_{p,l}^{T}\hat{\varphi}_{p,l}^{\prime}\right)\hat{V}_{p,0}^{T},\,\forall p\in\{1,\ldots,m\}.\label{eq:Phi_pj}
\end{eqnarray}
\begin{rem}
\label{rem:vanishing gradient}If $\Phi_{p}$ suffers from the vanishing
gradient problem, i.e., $\left\Vert \Xi_{l}\right\Vert _{F}\approx0$
for all $l\in\{p+1,\dots,m\}$, then $\Phi_{p}^{\prime}=\left(\stackrel{\curvearrowleft}{\stackrel[l=p+1]{m}{\prod}}\left(I_{n}+\Xi_{l}\right)\right)\Lambda_{p}\approx\Lambda_{p}$.
For an equivalent fully-connected DNN, i.e., in absence of shortcut
connections, $\left\Vert \Phi_{p}^{\prime}\right\Vert _{F}=\left\Vert \left(\stackrel{\curvearrowleft}{\stackrel[l=p+1]{m}{\prod}}\Xi_{l}\right)\Lambda_{p}\right\Vert \approx0$.
Thus, the shortcut connection circumvents the vanishing gradient problem
in the ResNet when $\Phi_{p}$ has a vanishing gradient.
\end{rem}
Based on the subsequent stability analysis, the control input is designed
as
\begin{eqnarray}
u & \triangleq & \dot{x}_{d}-\Phi^{\hat{\theta}}(x)-\sigma_{e}e-\sigma_{s}\text{sgn}(e),\label{eq:u}
\end{eqnarray}
where $\sigma_{e},\sigma_{s}\in\mathbb{R}_{>0}$ are constant control
gains, and $\text{sgn}(\cdot)$ denotes the vector signum function. 

\section{Stability Analysis}

To facilitate the subsequent analysis, let $z\triangleq[e^{T},\tilde{\theta}^{T}]^{T}\in\mathbb{R}^{\Psi}$
denote a concatenated state, where $\Psi\triangleq n+\Sigma_{p=1}^{m}\Sigma_{j=0}^{k_{p}}L_{p,j}L_{p,j+1}$.
Consider the candidate Lyapunov function $\mathcal{V}_{L}:\mathbb{R}^{\Psi}\rightarrow\mathbb{R}_{\ge0}$
defined as
\begin{eqnarray}
\mathcal{V}_{L}\left(z\right) & \triangleq & \frac{1}{2}e^{T}e+\frac{1}{2}\tilde{\theta}^{T}\Gamma^{-1}\tilde{\theta},\label{eq: Lyap function}
\end{eqnarray}
which satisfies the inequality $\alpha_{1}\left\Vert z\right\Vert ^{2}\le\mathcal{V}_{L}\left(z\right)\le\alpha_{2}\left\Vert z\right\Vert ^{2},$
where $\alpha_{1},\alpha_{2}\in\mathbb{R}_{>0}$ are known constants.
The universal function approximation property in (\ref{eq:UAP}) holds
only on the compact domain $\Omega$; hence, the subsequent stability
analysis requires ensuring $x(t)\in\Omega$ for all $t\in\mathbb{R}_{\geq0}$.
This is achieved by yielding a stability result which constrains $z$
in a compact domain. Consider the compact domain $\mathcal{D}\triangleq\{\varsigma\in\mathbb{R}^{\Psi}:\left\Vert \varsigma\right\Vert <\kappa\}$
in which $z$ is supposed to lie, where $\kappa\in\mathbb{R}_{>0}$
is a bounding constant. The subsequent analysis shows that $z(t)\in\mathcal{D}$
for all $t\in\mathbb{R}_{\geq0}$, if $z$ is initialized within the
set $\mathcal{S}\triangleq\{\varsigma\in\mathbb{R}^{\Psi}:\left\Vert \varsigma\right\Vert <\sqrt{\frac{\alpha_{1}}{\alpha_{2}}}\kappa\}$.

Taking the time-derivative of (\ref{eq: e}), substituting in (\ref{eq: xDot})
and (\ref{eq:u}), and substituting in (\ref{eq:UAP}) yields the
closed-loop error system
\begin{eqnarray}
\dot{e} & = & \Phi^{\theta^{*}}(x)-\Phi^{\hat{\theta}}(x)+\varepsilon(x)-\sigma_{e}e-\sigma_{s}\text{sgn}(e).\label{eq:edot_fs}
\end{eqnarray}
The ResNet in (\ref{eq:ResNet1_arch}) is nonlinear in terms of the
weights. Adaptive control design for nonlinearly parameterized systems
is known to be a difficult problem \cite{Annaswamy.Skantze.ea1998}.
A number of adaptive control methods have been developed to address
the challenges posed by a nonlinear parameterization \cite{Annaswamy.Skantze.ea1998,Kojic.Annaswamy.ea1999,Lin.Qian2002,Lin2002,Qu2006,Roy.Bhasin.ea2017b,Lewis1996a,Patil.Le.ea2022}.
In particular, first-order Taylor series approximation-based techniques
have shown promising results for neural network-based adaptive controllers
\cite{Lewis1996a,Ge2002,Patil.Le.ea2022}. Specifically, the result
in \cite{Patil.Le.ea2022} uses a first-order Taylor series approximation
to derive weight adaptation laws for a fully-connected DNN-based adaptive
controller. Thus, motivation exists to explore a Taylor series approximation-based
design to derive adaptation laws for the ResNet. For the ResNet in
(\ref{eq:ResNet1_arch}), a first-order Taylor series approximation-based
error model is given by \cite[Eq. 22]{Lewis1996b}
\begin{eqnarray}
\Phi^{\theta^{*}}(x)-\Phi^{\hat{\theta}}(x) & = & \Phi^{\prime}\tilde{\theta}+\mathcal{O}\left(\left\Vert \tilde{\theta}\right\Vert ^{2}\right),\label{eq:Taylor_approx}
\end{eqnarray}
where $\mathcal{O}\left(\left\Vert \tilde{\theta}\right\Vert ^{2}\right)$
denotes  higher-order terms. Since $\left\Vert x_{d}(t)\right\Vert \leq\overline{x_{d}}$
for all $t\in\mathbb{R}_{\geq0}$, $x$ can be bounded as $\left\Vert x\right\Vert \leq\left\Vert e+x_{d}\right\Vert \leq\left\Vert z\right\Vert +\left\Vert x_{d}\right\Vert \leq\kappa+\overline{x_{d}}$,
based on the definition of $\mathcal{D}$, when $z\in\mathcal{D}$.
Hence, since the ResNet is smooth, there exists a known constant $\overline{\Delta}\in\mathbb{R}_{>0}$
such that $\left\Vert \mathcal{O}\left(\left\Vert \tilde{\theta}\right\Vert ^{2}\right)\right\Vert \leq\overline{\Delta}$,
when $z\in\mathcal{D}$. Then, substituting (\ref{eq:Taylor_approx})
into (\ref{eq:edot_fs}), the closed-loop error system can be expressed
as 
\begin{eqnarray}
\dot{e} & = & \Phi^{\prime}\tilde{\theta}+\mathcal{O}\left(\left\Vert \tilde{\theta}\right\Vert ^{2}\right)+\varepsilon(x)-\sigma_{e}e-\sigma_{s}\text{sgn}(e).\label{eq:edot_fs-1}
\end{eqnarray}

Then, using (\ref{eq:adaptation_law}) and (\ref{eq:edot_fs-1}) yields
\begin{eqnarray}
\dot{z} & = & h(z,t),\label{eq:zdot}
\end{eqnarray}
where $h:\mathbb{R}^{\Psi}\times\mathbb{R}_{\geq0}\to\mathbb{R}^{\Psi}$
is defined as
\begin{equation}
h(z,t)\triangleq\left[\begin{array}{c}
\left(\begin{array}{c}
\Phi^{\prime}\tilde{\theta}+\mathcal{O}\left(\left\Vert \tilde{\theta}\right\Vert ^{2}\right)+\varepsilon(x)\\
\quad-\sigma_{e}e-\sigma_{s}\text{sgn}(e)
\end{array}\right)\\
-\Gamma\Phi^{\prime T}e
\end{array}\right].\label{eq:h_def}
\end{equation}
 Based on the nonsmooth analysis technique in \cite{Fischer.Kamalapurkar.ea2013},
the following theorem establishes the invariance properties of Filippov
solutions to (\ref{eq:zdot}) and provides guarantees of asymptotic
tracking error convergence for the system in (\ref{eq: xDot}). 
\begin{thm}
\label{thm:main theorem}For the dynamical system in (\ref{eq: xDot}),
the controller in (\ref{eq:u}) and the adaptation law in (\ref{eq:adaptation_law})
ensure asymptotic tracking error convergence in the sense that $z,u,\dot{\hat{\theta}}\in\mathcal{L}_{\infty}$
and $\underset{t\to\infty}{\lim}\left\Vert e(t)\right\Vert =0$, provided
Assumptions \ref{assm:UAP} and \ref{assm:theta bound} hold, $z(0)\in\mathcal{S}$,
and the following gain condition is satisfied:
\begin{eqnarray}
\sigma_{s} & > & \overline{\varepsilon}+\overline{\Delta}.\label{eq:gain_condition}
\end{eqnarray}
\end{thm}
\begin{IEEEproof}
Let $\partial\mathcal{V}_{L}$ denote the Clarke gradient of $\mathcal{V}_{L}$
defined in \cite[p. 39]{Clarke1990}. Since $z\mapsto\mathcal{V}_{L}(z)$
is continuously differentiable, $\partial\mathcal{V}_{L}(z)=\{\nabla\mathcal{V}_{L}(z)\}$,
where $\nabla$ denotes the standard gradient operator. Based on (\ref{eq:h_def})
and the chain rule in \cite[Thm 2.2]{Shevitz1994}, it can be verified
that $t\to\mathcal{V}_{L}(z(t))$ satisfies the differential inclusion
\begin{eqnarray}
\dot{\mathcal{V}}_{L} & \overset{a.a.t.}{\in} & \underset{\xi\in\partial\mathcal{V}_{L}\left(z\right)}{\bigcap}\xi^{T}K\left[h\right](z,t)\nonumber \\
 & = & \nabla\mathcal{V}_{L}\left(z\right)^{T}K\left[h\right](z,t)\nonumber \\
 & = & e^{T}\left(\mathcal{O}\left(\left\Vert \tilde{\theta}\right\Vert ^{2}\right)+\varepsilon(x)\right)-\sigma_{e}\left\Vert e\right\Vert ^{2}\nonumber \\
 &  & -\sigma_{s}e^{T}K\left[\text{sgn}\right](e)+e^{T}\Phi^{\prime}\tilde{\theta}-\tilde{\theta}^{T}\Phi^{\prime T}e,\label{eq:Lyap_deriv}
\end{eqnarray}
for all $z\in\mathcal{D}$. Using the fact that $e^{T}K\left[\text{sgn}\right](e)=\left\Vert e\right\Vert _{1}$,
(\ref{eq:Lyap_deriv}) can be bounded as
\begin{equation}
\dot{\mathcal{V}}_{L}\overset{a.a.t.}{\leq}-\sigma_{e}\left\Vert e\right\Vert ^{2}+e^{T}\left(\mathcal{O}\left(\left\Vert \tilde{\theta}\right\Vert ^{2}\right)+\varepsilon(x)\right)-\sigma_{s}\left\Vert e\right\Vert _{1},\label{eq:VLdot}
\end{equation}
for all $z\in\mathcal{D}$. Based on Holder's inequality, triangle
inequality, and the fact that $\left\Vert e\right\Vert \leq\left\Vert e\right\Vert _{1}$,
the following inequality can be obtained: $e^{T}\left(\mathcal{O}\left(\left\Vert \tilde{\theta}\right\Vert ^{2}\right)+\varepsilon(x)\right)$$\leq\left\Vert e\right\Vert _{1}\left(\left\Vert \mathcal{O}\left(\left\Vert \tilde{\theta}\right\Vert ^{2}\right)\right\Vert +\left\Vert \varepsilon(x)\right\Vert \right)\leq\left(\bar{\varepsilon}+\overline{\Delta}\right)\left\Vert e\right\Vert _{1}$.
Then, provided the gain condition in (\ref{eq:gain_condition}) is
satisfied, the right-hand side of (\ref{eq:VLdot}) can be upper-bounded
as
\begin{eqnarray}
\dot{\mathcal{V}}_{L} & \overset{a.a.t.}{\leq} & -\sigma_{e}\left\Vert e\right\Vert ^{2},\label{eq:VLdot_final}
\end{eqnarray}
for all $z\in\mathcal{D}$. Based on (\ref{eq:VLdot_final}), invoking
\cite[Corollary 1]{Fischer.Kamalapurkar.ea2013} yields $z\in\mathcal{L}_{\infty}$
and $\underset{t\to\infty}{\lim}\left\Vert e(t)\right\Vert =0$, when
$z\in\mathcal{D}$. Using (\ref{eq:VLdot_final}), $\alpha_{1}\left\Vert z(t)\right\Vert ^{2}\le\mathcal{V}_{L}\left(z(t)\right)\leq\mathcal{V}_{L}\left(z(0)\right)\le\alpha_{2}\left\Vert z(0)\right\Vert ^{2}$,
when $z(t)\in\mathcal{D}$. Thus, $\left\Vert z(t)\right\Vert <\sqrt{\frac{\alpha_{2}}{\alpha_{1}}}\left\Vert z(0)\right\Vert $,
when $z(t)\in\mathcal{D}$. Therefore, $z(t)\in\mathcal{D}$ is always
satisfied if $\left\Vert z(0)\right\Vert \leq\sqrt{\frac{\alpha_{1}}{\alpha_{2}}}\kappa$,
i.e., $z(0)\in\mathcal{S}$. To show $x\in\Omega$ for ensuring the
universal function approximation holds, consider the set $\Upsilon\subseteq\Omega$
defined as $\Upsilon\triangleq\{\zeta\in\Omega:\left\Vert \zeta\right\Vert \leq\kappa+\overline{x_{d}}\}$.
Since $\left\Vert z\right\Vert \leq\kappa$ implies $\left\Vert e\right\Vert \leq\kappa$,
the following relation holds: $\left\Vert x\right\Vert \leq\left\Vert e+x_{d}\right\Vert \leq\kappa+\overline{x_{d}}$.
Therefore, $x(t)\in\Upsilon\subseteq\Omega$ for all $t\in\mathbb{R}_{\geq0}$.
Additionally, due to the facts that $(x,\hat{\theta})\to\Phi^{\hat{\theta}}(x)$
is smooth, $x\in\Omega$, and $\hat{\theta}\in\mathcal{B}$, it follows
that $\Phi^{\hat{\theta}}(x)$ is bounded. Since each term on the
right-hand side of (\ref{eq:u}) is bounded, the control input $u\in\mathcal{L}_{\infty}$.
Since $\phi_{p,j}$ and $\phi_{p,j}^{\prime}$ are smooth for all
$(p,j)\in\{1,\ldots,m\}\times\{0,\ldots,k_{p}\}$, it follows from
(\ref{eq:Phi_prime})-(\ref{eq:Phi_pj}) that $\Phi^{\prime}$ is
bounded. Then, every term on the right-hand side of (\ref{eq:adaptation_law})
is bounded, and hence, $\dot{\hat{\theta}}$ is bounded.
\end{IEEEproof}
\begin{rem}
\label{rem:DCAL}If the ResNet is used to approximate the desired
drift $f(x_{d})$ instead of the actual drift $f(x)$, the control
design and analysis method in our preliminary work in \cite{Patil.Le.ea.2022}
can be used with the developed method to yield asymptotic tracking
error convergence for any value of the initial condition $e(0)\in\mathbb{R}^{n}$.
\end{rem}
\begin{rem}
\label{rem:sigma mod}If the sliding-mode term $\sigma_{s}\text{sgn}(e)$
is removed from the control input, the adaptation law in (\ref{eq:adaptation_law})
can be modified with standard robust modification techniques such
as sigma modification or e-modification \cite[Ch. 8]{Ioannou1996}
, where a uniformly ultimately bounded tracking result can be obtained
without requiring knowledge of the bounds $\bar{\theta}$, $\bar{\Delta}$,
and $\bar{\varepsilon}$. 
\end{rem}

\section{Simulations}

Monte Carlo simulations are provided to demonstrate the performance
of the developed ResNet-based adaptive controller, and the results
are compared with a fully-connected DNN-based adaptive controller
\cite{Patil.Le.ea2022}. The system in (\ref{eq: xDot}) is considered
with the state dimension $n=10$. The unknown drift vector field in
(\ref{eq: xDot}) is modeled as $f(x)=Ay(x)$, where $A\in\mathbb{R}^{n\times6n}$
is a random matrix with all elements belonging to the uniform random
distribution $U(0,0.1)$, and $y(x)\triangleq[x^{T},\tanh(x)^{T},\sin(x)^{T},\mathrm{sech}(x)^{T},(x\odot x)^{T},(x\odot x\odot x)^{T}]^{T}$,
where $\odot$ denotes the element-wise product operator. All elements
of the initial state $x(0)$ are selected from the distribution $U(0,2)$.
The reference trajectory is selected as $x_{d}(t)=\left[0.5+\sin(\omega_{1}t),\ldots,0.5+\sin(\omega_{n}t)\right]$,
where $\omega_{1},\ldots,\omega_{n}\sim U(0,20)$. The configuration
of the ResNet in (\ref{eq:ResNet1_arch}) is selected with 20 hidden
layers, a shortcut connection across each hidden layer, and 10 neurons
in each layer. The hyperbolic tangent activation function is used
in each node of the ResNet. The results are compared with an equivalent
fully-connected DNN-based adaptive control, i.e., the same configuration
as the ResNet but without shortcut connections. The control and adaptation
gains are selected as $\sigma_{e}=2$, $\sigma_{s}=2$, and $\Gamma=I_{\Sigma_{p=1}^{m}\Sigma_{j=0}^{k_{p}}L_{p,j}L_{p,j+1}}$.

The performance of both the ResNet and the fully-connected DNN-based
adaptive controller is sensitive to initial weights. To account for
the sensitivity of performance to weight initialization, the initial
weights for each method are obtained using a Monte Carlo method. In
the Monte Carlo method, 10,000 simulations are performed, where the
initial weights in each simulation are selected from $U(-0.05,0.05)$,
and the cost $J=\int_{0}^{10}\left(e^{T}(t)Qe(t)+u^{T}(t)Ru(t)\right)dt$
is evaluated in each simulation with $Q=I_{10}$ and $R=0.01I_{10}$.
For a fair comparison between the ResNet and the fully-connected DNN,
the simulation results yielding the least $J$ for each architecture
are compared.

\begin{table}
\caption{\label{tab:Performance-Comparison}Performance Comparison}

\centering{}%
\begin{tabular}{cccc}
\hline 
Architecture &
$\left\Vert e_{\mathrm{rms}}\right\Vert $ &
$\left\Vert \tilde{f}_{\mathrm{rms}}\right\Vert $ &
$\left\Vert u_{\mathrm{rms}}\right\Vert $\tabularnewline
\hline 
\noalign{\vskip5pt}
\vspace{5pt}ResNet &
0.329 &
3.395 &
24.332\tabularnewline
Fully-Connected &
0.912 &
9.636 &
24.816\tabularnewline
Shallow NN with 10 neurons &
0.727 &
6.187 &
24.681\tabularnewline
Shallow NN with 100 neurons &
0.560 &
4.979 &
24.662\tabularnewline
\hline 
\end{tabular}
\end{table}
\begin{figure}[h]
\begin{centering}
\includegraphics[width=9cm]{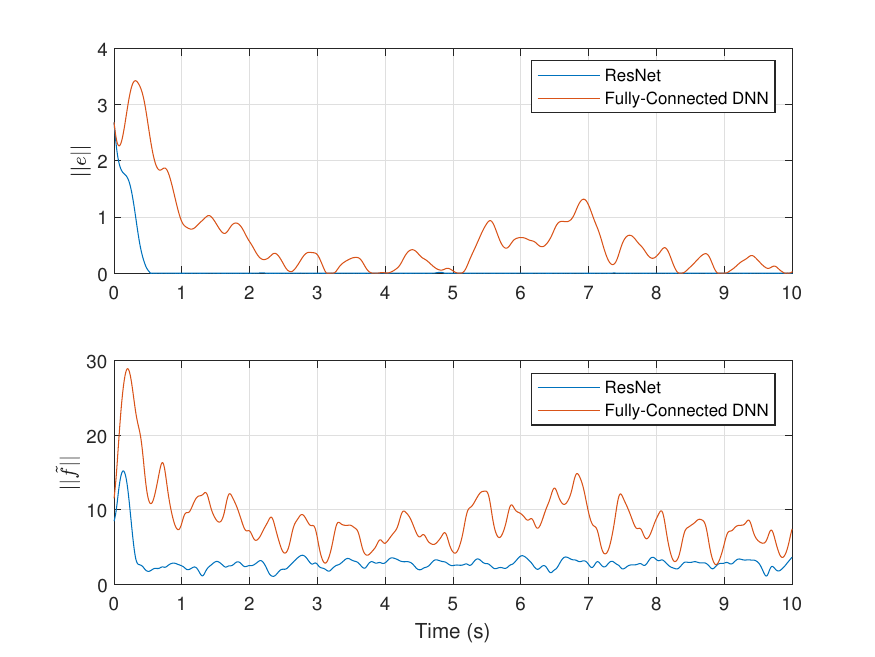}
\par\end{centering}
\caption{\label{fig:Error Plots}Plots of the tracking error norm $\left\Vert e\right\Vert $
and function approximation error norm $\left\Vert \tilde{f}\right\Vert $
with ResNet and fully-connected DNN-based adaptive controller.}
\end{figure}

\begin{figure}
\begin{centering}
\includegraphics[width=9cm]{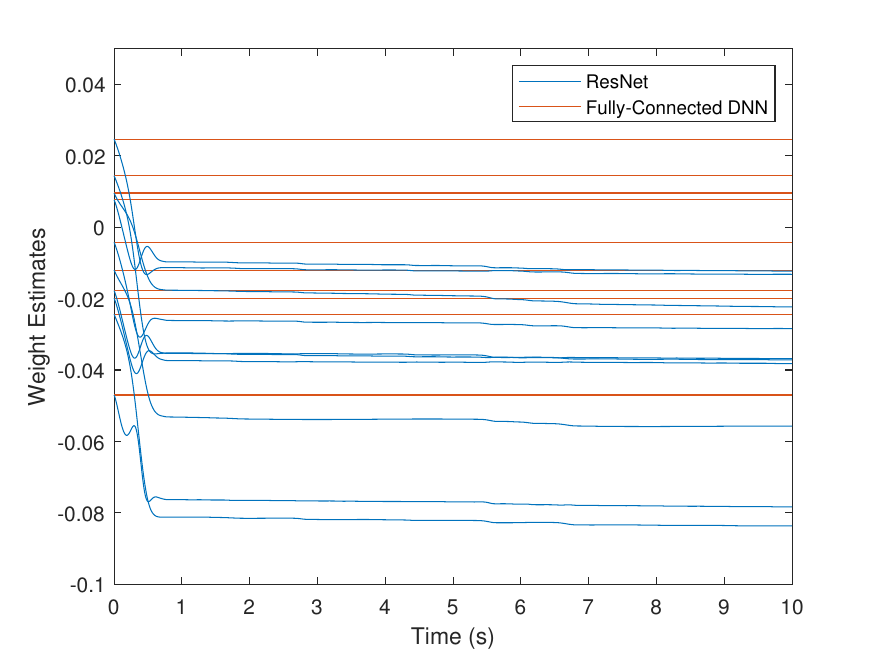}
\par\end{centering}
\caption{\label{fig:combined weight estimates}Plot of the weight estimates
of the ResNet and fully-connected DNN. There are a total of 2,000
individual weights in each architecture. For better visualization,
10 arbitrarily selected weights are shown. The fully-connected DNN
weights adapt slowly due to the problem of vanishing gradients. However,
the ResNet weights are able to adapt faster since the ResNet does
not have vanishing gradients.}
\end{figure}

\begin{figure}[H]
\centering{}\includegraphics[width=9cm]{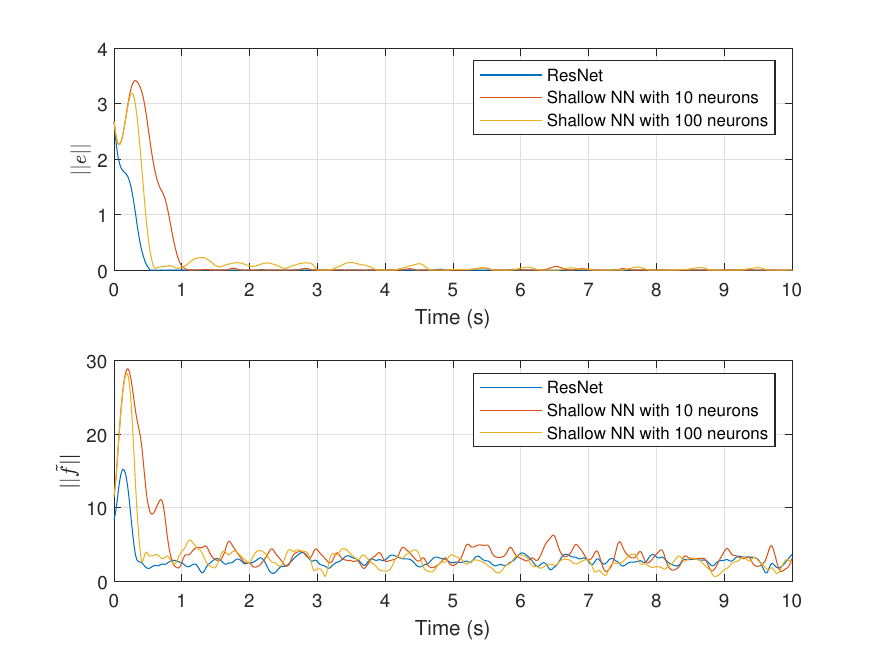}\caption{\textcolor{blue}{\label{fig:Shallow Comparison}}Comparative plots
of the tracking error norm $\left\Vert e\right\Vert $ and the function
approximation error norm $\left\Vert \tilde{f}\right\Vert $ with
the ResNet and a shallow NN with 10 neurons.}
\end{figure}

\begin{figure}[H]
\begin{centering}
\includegraphics[width=9cm]{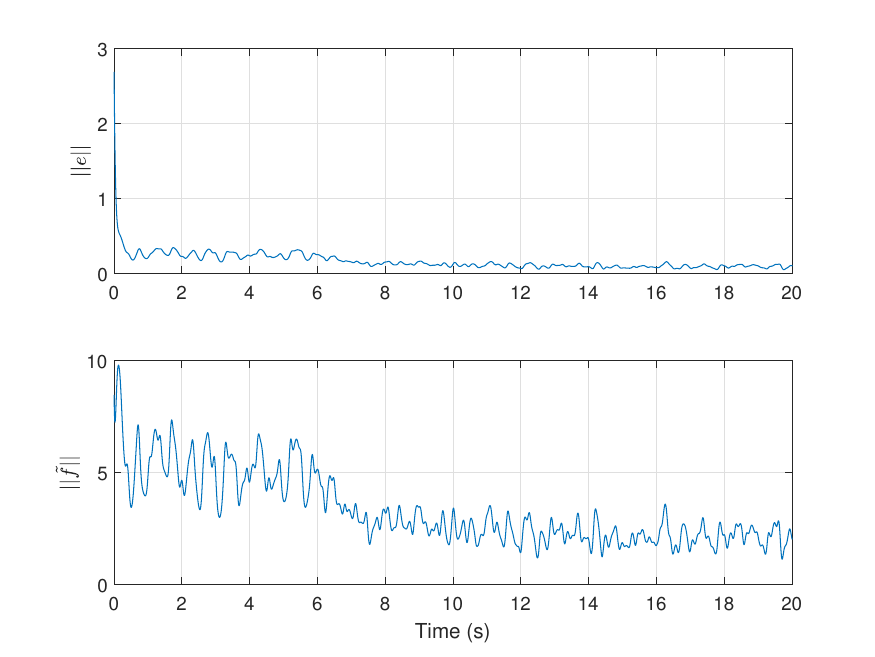}
\par\end{centering}
\caption{\label{fig:sigma mod}Plots demonstrating the tracking and function
approximation errors with the ResNet-based controller, using sigma
modification in the adaptation law.}
\end{figure}

Table \ref{tab:Performance-Comparison} provides the norm of the root
mean square (RMS) tracking error, function approximation error, and
control input given by $\left\Vert e_{\mathrm{rms}}\right\Vert $,
$\left\Vert \tilde{f}_{\mathrm{rms}}\right\Vert $, and $\left\Vert u_{\mathrm{rms}}\right\Vert $,
respectively. In comparison to the fully-connected DNN, the ResNet
shows 63.93\% and 64.77\% decrease in the norms of the tracking and
function approximation errors, respectively. As shown in Figure \ref{fig:Error Plots},
the fully-connected DNN exhibits a comparatively poor tracking and
function approximation performance. As mentioned in Remark \ref{rem:vanishing gradient},
fully-connected DNNs suffer from the vanishing gradient problem. Thus,
the fully-connected DNN weights remain approximately constant as shown
in Figure \ref{fig:combined weight estimates}. Consequently, the
fully-connected DNN-based feedforward term fails to compensate for
the uncertainty in the system which yields a relatively poor tracking
and function approximation. In contrast to the fully-connected DNN,
the presence of shortcut connections in the ResNet eliminates the
vanishing gradient problem as mentioned in Remark \ref{rem:vanishing gradient}.
As a result, the ResNet weights are able to compensate for the system
uncertainty as shown in Figure \ref{fig:combined weight estimates}
which yields improved tracking and function approximation performance.
Additionally, the ResNet requires approximately the same control effort
as the fully-connected DNN. Therefore, the ResNet improves the tracking
performance without requiring a higher control effort in comparison
to the fully-connected DNN.
Additionally, to demonstrate performance comparison of the ResNet
with shallow NNs, comparative simulations two different configurations
are used for the shallow NN architecture given by $V_{1}^{T}\phi(V_{0}^{T}x)$.
In the first configuration, 10 neurons are used in the hidden layer
of the shallow NN, i.e., the same number of neurons as in hidden layer
of the ResNet. In the second configuration, we use 100 neurons in
the hidden layer of the shallow NN, which yields the same total number
of individual weights as the ResNet (i.e., 2000). As evident from
Figure \ref{fig:Shallow Comparison}, the ResNet significantly outperforms
both of the shallow NN configurations. Both of the shallow NN configurations
exhibit overshoot in the tracking and function approximation errors,
unlike the ResNet. The ResNet achieves rapid tracking error convergence
in approximately 0.5 seconds. In comparison, both of the shallow NN
configurations fail to demonstrate exact tracking error convergence,
despite containing the same robust state-feedback gains as the ResNet-based
controller.\textcolor{blue}{}
\textcolor{blue}{{} }Note that in practice, one can tune a ResNet to
contain sufficiently large number of layers, neurons, and shortcut
connections, which would yield a small value of the approximation
error $\varepsilon$, which can be easily compensated using the sliding-mode
term to yield asymptotic tracking. Additionally, to avoid using a
sliding-mode term or the knowledge of the bound on $\theta^{*}$,
other standard robust modification techniques such as sigma modification
or e-modification \cite[Ch. 8]{Ioannou1996} can be used in the adaptation
law, where a uniformly ultimately bounded (UUB) tracking result can
be obtained, as stated in Remark \ref{rem:sigma mod}. The sliding-mode
term is used to show an asymptotic tracking result, but it is not
central to the main development. Figure \ref{fig:sigma mod} demonstrates
the simulation results where the sliding-mode term $\sigma_{s}\mathrm{sgn}(e)$
is omitted and the e-modification based update law $\dot{\hat{\theta}}=-\sigma_{\theta}\left\Vert e\right\Vert \hat{\theta}+\Gamma\Phi^{\prime T}e$
is used with $\sigma_{\theta}=1$ and $\sigma_{e}=20$. As evident,
the ResNet-based controller with the e-modification is able to track
the desired trajectory with an ultimate bound of 0.15 on the tracking
error, and root mean square (RMS) values of 0.269, 4.669, and 23.194
for the tracking error, function approximation error, and control
input norms, respectively. 

\section{Conclusion and Future Work}

This paper provides the first result on Lyapunov-derived adaptation
laws for the weights of each layer of a ResNet-based adaptive controller.
A nonsmooth Lyapunov-based analysis is provided to guarantee  asymptotic
tracking error convergence. Comparative Monte Carlo simulations are
provided to demonstrate the performance of the developed ResNet-based
adaptive controller. The developed ResNet-based adaptive controller
provides approximately 64\% improvement in the  tracking and function
approximation performance, in comparison to an equivalent fully-connected
DNN-based adaptive controller. Additionally, the ResNet overcomes
the vanishing gradient problem present in the fully-connected DNN.

Future work can explore incorporating a long short-term memory component
in the ResNet architecture, based on our recent work \cite{Griffis.Patil.ea23_2},
to model uncertainties with long-term temporal dependencies. Additionally,
composite adaptive methods can be explored that incorporate a prediction
error of the uncertainty, in addition to the tracking error, in the
adaptation law.

\bibliographystyle{ieeetr}
\bibliography{encr,master,ncr}

\begin{thebibliography}{10}

\bibitem{Kidger.Lyons2020}
P.~Kidger and T.~Lyons, ``Universal approximation with deep narrow networks,''
  in {\em Conf. Learn. Theory}, pp.~2306--2327, 2020.

\bibitem{Brunton.Kutz2019}
S.~L. Brunton and J.~N. Kutz, {\em Data-driven science and engineering: Machine
  learning, dynamical systems, and control}.
\newblock Cambridge University Press, 2019.

\bibitem{Lewis1996a}
F.~Lewis, A.~Yesildirek, and K.~Liu, ``Multilayer neural net robot controller:
  structure and stability proofs,'' {\em IEEE Trans. Neural Netw.}, vol.~7,
  no.~2, pp.~388--399, 1996.

\bibitem{Lewis1999a}
F.~L. Lewis, S.~Jagannathan, and A.~Yesildirak, {\em Neural network control of
  robot manipulators and nonlinear systems}.
\newblock Philadelphia, PA: CRC Press, 1998.

\bibitem{Ge2002}
S.~S. Ge, C.~C. Hang, T.~H. Lee, and T.~Zhang, {\em Stable Adaptive Neural
  Network Control}.
\newblock Boston, MA: Kluwer Academic Publishers, 2002.

\bibitem{Patre2008}
P.~M. Patre, W.~MacKunis, K.~Kaiser, and W.~E. Dixon, ``Asymptotic tracking for
  uncertain dynamic systems via a multilayer neural network feedforward and
  {RISE} feedback control structure,'' {\em IEEE Trans. Autom. Control},
  vol.~53, no.~9, pp.~2180--2185, 2008.

\bibitem{Patre2010a}
P.~Patre, S.~Bhasin, Z.~D. Wilcox, and W.~E. Dixon, ``Composite adaptation for
  neural network-based controllers,'' {\em IEEE Trans. Autom. Control},
  vol.~55, no.~4, pp.~944--950, 2010.

\bibitem{Sun.Greene.ea2021}
R.~Sun, M.~Greene, D.~Le, Z.~Bell, G.~Chowdhary, and W.~E. Dixon,
  ``Lyapunov-based real-time and iterative adjustment of deep neural
  networks,'' {\em IEEE Control Syst. Lett.}, vol.~6, pp.~193--198, 2022.

\bibitem{Joshi.Chowdhary2019}
G.~Joshi and G.~Chowdhary, ``Deep model reference adaptive control,'' in {\em
  Proc. IEEE Conf. Decis. Control}, pp.~4601--4608, 2019.

\bibitem{Joshi.Virdi.ea2020a}
G.~Joshi, J.~Virdi, and G.~Chowdhary, ``Asynchronous deep model reference
  adaptive control,'' in {\em Conf. Robot Learn.}, 2020.

\bibitem{Le.Greene.ea2021}
D.~Le, M.~Greene, W.~Makumi, and W.~E. Dixon, ``Real-time modular deep neural
  network-based adaptive control of nonlinear systems,'' {\em IEEE Control
  Syst. Lett.}, vol.~6, pp.~476--481, 2022.

\bibitem{Patil.Le.ea2022}
O.~Patil, D.~Le, M.~Greene, and W.~E. Dixon, ``Lyapunov-derived control and
  adaptive update laws for inner and outer layer weights of a deep neural
  network,'' {\em IEEE Control Syst Lett.}, vol.~6, pp.~1855--1860, 2022.

\bibitem{Goodfellow2016}
I.~Goodfellow, Y.~Bengio, A.~Courville, and Y.~Bengio, {\em Deep {L}earning},
  vol.~1.
\newblock MIT press Cambridge, 2016.

\bibitem{He.Zhang.ea2016}
K.~He, X.~Zhang, S.~Ren, and J.~Sun, ``Deep residual learning for image
  recognition,'' in {\em Proc. IEEE Conf. Comput. Vis. Pattern Recognit.},
  pp.~770--778, 2016.

\bibitem{Hardt.Ma2017}
M.~Hardt and T.~Ma, ``Identity matters in deep learning,'' {\em Int. Conf.
  Learn. Represent.}, 2017.

\bibitem{Nar.Sastry2018}
K.~Nar and S.~Sastry, ``Residual networks: Lyapunov stability and convex
  decomposition,'' {\em arXiv preprint arXiv:1803.08203}, 2018.

\bibitem{Tai.Yang.ea2017}
Y.~Tai, J.~Yang, and X.~Liu, ``Image super-resolution via deep recursive
  residual network,'' in {\em Proc. IEEE Conf. Comput. Vis. Pattern Recognit.},
  pp.~3147--3155, 2017.

\bibitem{Li.Fang.ea2018}
J.~Li, F.~Fang, K.~Mei, and G.~Zhang, ``Multi-scale residual network for image
  super-resolution,'' in {\em Proc. Eur. Conf. Comput. Vis.}, pp.~517--532,
  2018.

\bibitem{Boroumand.Chen.ea2018}
M.~Boroumand, M.~Chen, and J.~Fridrich, ``Deep residual network for
  steganalysis of digital images,'' {\em IEEE Trans. Inf. Forensics Secur.},
  vol.~14, no.~5, pp.~1181--1193, 2018.

\bibitem{Tan.Qian.ea2018}
T.~Tan, Y.~Qian, H.~Hu, Y.~Zhou, W.~Ding, and K.~Yu, ``Adaptive very deep
  convolutional residual network for noise robust speech recognition,'' {\em
  IEEE/ACM Trans. Audio, Speech, Language Process.}, vol.~26, no.~8,
  pp.~1393--1405, 2018.

\bibitem{Patil.Le.ea.2022}
O.~S. Patil, D.~M. Le, E.~Griffis, and W.~E. Dixon, ``Deep residual neural
  network {(ResNet)}-based adaptive control: A {L}yapunov-based approach,'' in
  {\em Proc. IEEE Conf. Decis. Control}, pp.~3487--3492, 2022.

\bibitem{Paden1987}
B.~E. Paden and S.~S. Sastry, ``A calculus for computing {F}ilippov's
  differential inclusion with application to the variable structure control of
  robot manipulators,'' {\em IEEE Trans. Circuits Syst.}, vol.~34, pp.~73--82,
  Jan. 1987.

\bibitem{Bernstein2009}
D.~S. Bernstein, {\em Matrix {M}athematics}.
\newblock Princeton university press, 2009.

\bibitem{Lin.Jegelka2018}
H.~Lin and S.~Jegelka, ``Res{N}et with one-neuron hidden layers is a universal
  approximator,'' {\em Adv. Neural Inf. Process. Syst.}, vol.~31, 2018.

\bibitem{Tabuada.Gharesifard2020a}
P.~Tabuada and B.~Gharesifard, ``Universal approximation power of deep residual
  neural networks via nonlinear control theory,'' in {\em Int. Conf. Learn.
  Represent.}, 2020.

\bibitem{Lewis1996b}
F.~L. Lewis, A.~Yegildirek, and K.~Liu, ``Multilayer neural-net robot
  controller with guaranteed tracking performance,'' {\em IEEE Trans. Neural
  Netw.}, vol.~7, pp.~388--399, Mar. 1996.

\bibitem{Annaswamy.Skantze.ea1998}
A.~M. Annaswamy, F.~P. Skantze, and A.-P. Loh, ``Adaptive control of continuous
  time systems with convex/concave parametrization,'' {\em Automatica},
  vol.~34, no.~1, pp.~33--49, 1998.

\bibitem{Kojic.Annaswamy.ea1999}
A.~Koji{\'c}, A.~M. Annaswamy, A.-P. Loh, and R.~Lozano, ``Adaptive control of
  a class of nonlinear systems with convex/concave parameterization,'' {\em
  Syst. Control Lett.}, vol.~37, no.~5, pp.~267--274, 1999.

\bibitem{Lin.Qian2002}
W.~Lin and C.~Qian, ``Adaptive control of nonlinearly parameterized systems:
  the smooth feedback case,'' {\em IEEE Trans. Autom. Control}, vol.~47, no.~8,
  pp.~1249--1266, 2002.

\bibitem{Lin2002}
W.~Lin and C.~Qian, ``Adaptive control of nonlinearly parameterized systems: a
  nonsmooth feedback framework,'' {\em IEEE Trans. Autom. Control}, vol.~47,
  pp.~757--774, May 2002.

\bibitem{Qu2006}
Z.~Qu, R.~A. Hull, and J.~Wang, ``Globally stabilizing adaptive control design
  for nonlinearly-parameterized systems,'' {\em IEEE Trans. Autom. Control},
  vol.~51, pp.~1073--1079, June 2006.

\bibitem{Roy.Bhasin.ea2017b}
S.~B. Roy, S.~Bhasin, and I.~N. Kar, ``Robust gradient-based adaptive control
  of nonlinearly parametrized plants,'' {\em IEEE Control Syst. Lett.}, vol.~1,
  no.~2, pp.~352--357, 2017.

\bibitem{Fischer.Kamalapurkar.ea2013}
N.~Fischer, R.~Kamalapurkar, and W.~E. Dixon, ``La{S}alle-{Y}oshizawa
  corollaries for nonsmooth systems,'' {\em IEEE Trans. Autom. Control},
  vol.~58, pp.~2333--2338, Sep. 2013.

\bibitem{Clarke1990}
F.~H. Clarke, {\em Optimization and nonsmooth analysis}.
\newblock SIAM, 1990.

\bibitem{Shevitz1994}
D.~Shevitz and B.~Paden, ``Lyapunov stability theory of nonsmooth systems,''
  {\em IEEE Trans. Autom. Control}, vol.~39 no. 9, pp.~1910--1914, 1994.

\bibitem{Ioannou1996}
P.~Ioannou and J.~Sun, {\em Robust Adaptive Control}.
\newblock Prentice Hall, 1996.

\bibitem{Griffis.Patil.ea23_2}
E.~Griffis, O.~Patil, Z.~Bell, and W.~E. Dixon, ``Lyapunov-based long
  short-term memory ({L}b-{LSTM}) neural network-based control,'' {\em IEEE
  Control Syst. Lett.}, vol.~7, pp.~2976--2981, 2023.

\end{thebibliography}

\end{document}